\documentclass[12pt]{mn2e}

\usepackage{graphics,amssymb}

\title[Analytical approximations to theoretical Emds]{Analytical approximations to numerical solutions of theoretical emission measure distributions} 
\author[C. Jordan, J.-U. Ness, S. A. Sim]{C. Jordan$^1$\thanks{E-mail: 
cj@thphys.ox.ac.uk}, J.-U. Ness$^2$ and S. A. Sim$^3$\\
$^1$Department of Physics, Rudolf Peierls Centre for Theoretical
Physics, University of Oxford, 1 Keble Road, Oxford, OX1\,3NP, UK\\
$^2$ XMM-Newton Science Operations Centre, ESA, P.O. Box 78, 28691, 
Villanueva de la Ca\~nada, Madrid, Spain \\
$^3$ Research School of Astronomy and Astrophysics, Mount Stromlo Observatory,
Cotter Road, Weston Creek, ACT 2611, Australia} 

\begin{document}

\date{Accepted ; Received \today}

\pagerange{\pageref{1}--\pageref{9}} \pubyear{2011}

\maketitle

\label{1}

\begin{abstract} 
Emission line fluxes from cool stars are widely used to establish an
apparent emission measure distribution, $Emd_{\rm app}(T_{\rm e})$, between 
temperatures characteristic of the low transition region and the low corona. 
The true emission measure distribution, $Emd_{\rm t}(T_{\rm e})$, is 
determined by the energy balance and geometry adopted and, with a numerical 
model, can be used to predict $Emd_{\rm app}(T_{\rm e})$, to 
guide further modelling. The scaling laws that exist between coronal 
parameters arise from the dimensions of the terms in the energy balance 
equation. Here, analytical approximations to numerical solutions for 
$Emd_{\rm t}(T_{\rm e})$ are presented, which show how the constants in the 
coronal scaling laws are determined. The apparent emission measure 
distributions show a minimum value at some $T_{\rm o}$ and a maximum at the 
mean coronal temperature $T_{\rm c}$ (although in some stars, emission from 
active regions can contribute). It is shown that, for the energy balance and 
geometry adopted, the analytical values of the emission measure and electron
pressure at
$T_{\rm o}$ and $T_{\rm c}$, depend on only three parameters: the stellar 
surface gravity and the values of $T_{\rm o}$ and $T_{\rm c}$. The results 
are tested against full numerical solutions for $\epsilon$~Eri (K2~V) and are
applied to Procyon ($\alpha$ CMi; F5 IV/V). The analytical approximations can
be used to restrict the required range of full numerical solutions, to check 
the assumed geometry and to show where the adopted energy balance may not be 
appropriate.
   
\end{abstract}


\section{Introduction}
The observed fluxes of emission lines from cool stars, including the Sun, 
are now routinely used to derive apparent emission measures ($Em_{\rm app}$) 
for given lines and apparent emission measure distributions 
[$Emd_{\rm app}(T_{\rm e})$], by using different lines formed over a range of 
electron 
temperatures ($T_{\rm e}$). The precise definition of the emission measure 
differs between authors and that adopted here is given in Section 2.

When the {\it International Ultraviolet Explorer} ({\it IUE}) was operating, 
the $Emd_{\rm app}(T_{\rm e})$ of cool stars was known only between 
$T_{\rm e}$ $\simeq 10^4$~K and $\simeq 2 \times 10^5$~K (see e.g. Jordan et 
al. 1987). Early X-ray satellites and the {\it Extreme Ultraviolet Explorer} 
({\it EUVE}) together provided information above $\simeq 8 \times 10^5$~K, 
and under favourable circumstances the {\it EUVE} could detect a few lines
formed around $3 \times 10^5$~K (see e.g. Drake, Laming \& Widing 1995). 
The Goddard High Resolution 
Spectrograph (GHRS) and the Space Telescope Imaging Spectrograph (STIS) on 
the {\it Hubble Space Telescope} ({\it HST}) have given improved spectral 
resolution and sensitivity in the ultraviolet regions (e.g. Wood et al. 1996). 
In particular, in addition to lines formed in the low to mid transition region 
(at $10^4$ -- $2 \times 10^5$~K), the spectrum of an M-dwarf flare, obtained 
with the GHRS, showed the forbidden line of Fe\,{\sc xxi} at 1354~\AA\ (Maran
 et al. 1994). Also, spectra of G/K dwarfs obtained with the STIS showed 
the forbidden lines of Fe\,{\sc xii} at 1242 and 1349~\AA\  (Jordan et al. 
2001), formed in their upper transition region or inner corona.
The {\it Far Ultraviolet Spectroscopic Explorer} ({\it FUSE}) has observed 
the resonance lines of O\,{\sc vi} (formed around $3 \times 10^5$~K), a 
number of lines that are formed below 
$10^5$~K and further forbidden lines of Fe\,{\sc xviii} and Fe\,{\sc xix} 
(Redfield et al. 2003). Both {\it Chandra} and {\it XMM-Newton} now provide 
extensive information from lines formed at $T_{\rm e} \ge 8 \times 10^5$~K 
(see e.g. Sanz-Forcada, Favata \& Micela 2004). Thus the 
$Emd_{\rm app}(T_{\rm e})$ of 
a range of stars is now reasonably well constrained over all the temperature 
range of interest, and in some stars the observations of forbidden lines of 
highly ionized iron with the STIS and {\it FUSE} provide simultaneous 
measurements in the lower and upper transition region/corona. 

The true emission measure distribution [$Emd_{\rm t}(T_{\rm e})$] is 
{\it determined} 
by the energy balance in the outer atmosphere and the actual geometry.
By making numerical calculations in a chosen geometry, theoretical values of 
both $Emd_{\rm t}(T_{\rm e})$ and $Emd_{\rm app}(T_{\rm e})$ can be predicted 
and the latter can then be
 compared with that observed. In an example of this approach, Sim \& Jordan 
(2003a) used {\it EUVE} observations of $\epsilon$~Eri (HD22049, K2~V) to 
determine $Emd_{\rm app}(T_{\rm e})$. The numerical 
model assumed that in the upper transition region there is an energy balance 
between the divergence of the conductive flux and the radiation losses
and, as a boundary condition, was 
required to match the observed minimum in $Emd_{\rm app}(T_{\rm e})$. A 
spherically symmetric geometry was adopted and a non-thermal pressure term was 
included in the equation of hydrostatic equilibrium. The observed 
$Emd_{\rm app}(T_{\rm e})$ (and observed electron densities, 
$N_{\rm e}$) could be fitted by the values from the model provided the area 
filling factor
was $\simeq 1$ in the inner corona, and $\simeq 0.2$ in the mid transition
region. More recently, Ness \& Jordan (2008) have included an analysis of 
the X-ray spectrum obtained with the Low Energy Transmission Grating 
Spectrometer (LETGS) on {\it Chandra} to obtain an 
improved observed $Emd_{\rm app}(T_{\rm e})$ and also the relative element 
abundances. 
The atomic data from CHIANTI (v 5.2) (Landi et al. 2006) were used to update 
the values of the electron pressure ($P_{\rm e}$).
 Improved area factors, which were slightly smaller, 
were also found by iterating the first solution with a variable area factor. 

Early work on the implied area of emitting material in the solar transition 
region was carried out by Kopp (1972). Also, Torriccelli-Ciamponi, Einaudi \& 
Chiuderi (1982) used the presence of a minimum in the 
$Emd_{\rm app}(T_{\rm e})$ to 
constrain models for the heating of solar loop structures, in an extension of 
the approach used by Rosner, Tucker \& Vaiana (1978).  

Hearn (1975, 1977) derived scaling relations for the energy losses by 
conduction and radiation from coronae, by assuming minimum energy loss from 
a corona. His approach has some similarities to that adopted below and we 
compare his predictions with our results in Section 2.    

The results of a range of numerical models made for $\epsilon$~Eri by Ness \& 
Jordan (2008) showed 
that, provided the heating function can be expressed as an energy flux, 
there exist scaling laws between the calculated coronal emission measures, 
electron pressures, temperatures and the value of the stellar gravity, 
$g(r_{\rm c})$,
 at the radial distance at which the coronal temperature, $T_{\rm c}$, is 
reached. This is to be 
expected, since these scaling laws can be derived from simple dimensional 
arguments, with the constants being determined by the specific assumptions 
made. However, in numerical calculations in other than 
plane parallel geometry, and in calculations that include the non-thermal 
pressure term, it is difficult to see exactly what determines the constants.

Here we adopt some simplifications regarding the variation of $P_{\rm e}$ and 
the geometry, to find which chosen parameters control the full solutions to 
the $Emd_{\rm t}(T_{\rm e})$. It is shown that the analytical solutions given
 below depend only on the stellar surface gravity ($g_{*}$) and the boundary 
temperatures, 
$T_{\rm c}$ and $T_{\rm o}$. The values of $P_{\rm o}$, $P_{\rm c}$, 
$Em_{\rm t}(T_{\rm o})$ and $Em_{\rm t}(T_{\rm c})$ are all determined by the 
choice of $g_*$, $T_{\rm o}$ and $T_{\rm c}$. 

The simple approach adopted is also useful as a starting point for testing the 
following questions: does the emission have to come from some fraction of 
the surface area; is local deposition of a heating flux (other than from 
conduction) required below the 
corona; is a mean corona, rather than closed magnetic structures (with heights 
smaller than the isothermal pressure-squared scale-height) appropriate?

Although $T_{\rm o}$ is 
similar in the main-sequence stars studied ($\simeq 2 \times 10^5$~K), it is 
significantly higher in the F-star Procyon (HD61421, $\alpha$~CMi F5~IV/V) and
 other evolved stars, 
including single giants and RS~CVn binaries (as pointed out for Capella by 
Dupree et al. 1993). The giant stars and RS~CVn binaries are not considered 
here because we assume a plane parallel geometry, which becomes inappropriate
in lower gravity stars and in the coronae of close binaries. Also, 
consideration of their escape velocities
 suggests that the highest temperature material is likely to be magnetically 
confined, rather than occurring in a quiescent corona with a large scale 
height.     

The simple theoretical model adopted for stars with a quiescent corona is set
 out in Section 2, together with the analytical relations derived. Comparisons 
between the simple analytical results and those derived from the full 
computational models for $\epsilon$~Eri, by Ness \& Jordan (2008), are made 
in Section 3. Procyon is used as a further example in Section 4. The 
conclusions are discussed and summarized in Section 5.   

\section{Theory: stars with a quiescent corona}
In the Sun, the maximum in the observed $Emd_{\rm app}(T_{\rm e})$ occurs at 
the average temperature of the inner quiescent corona 
($T_{\rm c} \simeq 1.6 \times 10^6$~K). Except in observations at or above the
limb, the emission lines are formed predominantly over the first 
pressure-squared isothermal scale height, since the lines are collisionally 
excited, with fluxes $\propto N_{\rm e}^2$ (See Section 2.2.) As apparent in 
movies made using {\it Yohkoh}
data, the emission from the inner corona, away from active regions, shows 
little inhomogeneity and very few variations with time. If active regions are 
present, the emitting material is observed as a gradual decrease in the 
$Emd_{\rm app}(T_{\rm e})$ at temperatures larger than $T_{\rm c}$. In very 
early spatially unresolved studies, Neupert (1965) observed the variation with
 time of lines formed at different values of $T_{\rm e}$, over more than one 
solar rotation period. These support the above picture.   

In other main-sequence stars we expect a quiescent corona to be present, as 
well as active regions. $T_{\rm c}$ is assumed to correspond to 
the temperature at which the observed $Emd_{\rm app}(T_{\rm e})$ has its 
maximum value. At worst, this assumption gives an upper limit to $T_{\rm c}$. 
Ideally, the 
contribution of active regions should be studied through rotational modulation
 of high temperature X-ray lines, but this is currently difficult, owing to 
the large amount of observing time required. 

\subsection{The upper transition region}

The full numerical calculations were made using a spherically symmetric
atmosphere and included a non-thermal pressure term in the equation of 
hydrostatic equilibrium. The relevant equations have been given in Sim \& 
Jordan (2003a) and/or Griffiths \& Jordan (1998).
In the analytical results below, we adopt a plane parallel geometry and a 
constant emitting area. In Section 3 we compare the analytical results with 
those from the full spherically symmetric numerical models calculated by 
Ness \& Jordan (2008).    
 
The true emission measure for a given line is defined as
\begin{equation}
Em_{\rm t}(0.3) = \int_{\delta h} N_{\rm H} N_{\rm e} {\rm d}h~,      % (1)
\end{equation}
where $h$ is the height, $\delta h$ corresponds to the region over 
which $\log T_{\rm e}$ changes by $\pm 0.15$ about the optimum temperature of 
formation for the line and $N_{\rm H}$ is the number density of hydrogen.

In plane parallel geometry, the apparent emission measure for an optically 
thin line is given by
\begin{equation}
Em_{\rm app}(0.3) = \frac{1}{2} \int_{\delta h} N_{\rm H} N_{\rm e} {\rm d} 
h~, % (2)
\end{equation} 
where the factor of $1/2$ allows for the photons emitted in the outwards
direction. 

Eqn. (1) can also be expressed as 
\begin{equation}
Em_{\rm t}(0.3) = \frac{0.86 P_{\rm e}^2}{\sqrt {2} T_{\rm e}}\frac{{\rm d}h}
{{\rm d}T_{\rm e}}~, % (3) 
\end{equation}
where the electron pressure is defined as $P_{\rm e} = N_{\rm e} T_{\rm e}$,
and the factor of 0.86 arises from $N_{\rm H} \simeq 0.86 N_{\rm e}$. Here it
 has been assumed that $P_{\rm e}^2$ and ${\rm d}h/{\rm d}T_{\rm e}$ are 
constant over the region in which an individual line is formed. The
former is a good assumption, but the latter can initially be less accurate. 
However, in
actual calculations of line fluxes using a theoretical model, any variation of 
${\rm d}h/{\rm d}T_{\rm e}$ over the region of line formation is taken into 
account. The {\it starting values} of $Em_{\rm app}(0.3)$ are used to find
the initial $Emd_{\rm app}(0.3)(T_{\rm e})$, but following the calculation of
 the line fluxes, including the {\it full} contribution function for each 
line, $Emd_{\rm app}(0.3)(T_{\rm e})$ is optimised. Relative element 
abundances are
 also adjusted during this process. The details are given in Ness \& Jordan 
(2008). Note that eqn. (3) gives an expression for the true emission measure
 for a line formed at a given $T_{\rm e}$. Eqn. (3) is also 
used to define the true emission measure distribution, 
$Emd_{\rm t}(0.3)(T_{\rm e})$, since the quantities on the right-hand side are
all differentiable functions of $T_{\rm e}$. For simplicity 
of presentation, the label 0.3 is omitted in the equations below.  

The energy flux carried by thermal conduction, $F_{\rm c}(T_{\rm e})$, is 
given by 
\begin{equation}
F_{\rm c}(T_{\rm e}) = - \kappa T_{\rm e}^{5/2} 
                       \frac{{\rm d}T_{\rm e}}{{\rm d}h}~,    % (4)
\end{equation} 
where $\kappa T_{\rm e}^{5/2}$ is the coefficient of thermal conduction and,
from Spitzer (1956), $\kappa$ is taken to be $1.1 \times 10^{-6}$~
erg~cm$^{-1}$~s$^{-1}$~K$^{-7/2}$. Here, as in our full numerical
 calculations, we ignore the small variation in $\kappa$ with $N_{\rm e}$
and $T_{\rm e}$ (Spitzer 1956). In $\epsilon$~Eri, this amounts to
only 30 per cent between $\log T_{\rm e}$ = 5.3 and 6.5. There will be a 
difference between $F_{\rm c}(T_{\rm e})$ in the analytical and numerical, 
spherically symmetric, solutions, but given the limited extent of the region 
considered, this is not expected to be large in dwarf stars [see comments 
after eqn. (12)].

Eqn. (3) allows $F_{\rm c}(T_{\rm e})$ to be expressed in terms of 
$Em_{\rm t}$. I.e., 
\begin{equation}
F_{\rm c}(T_{\rm e}) = - \frac{0.86 \kappa }{\sqrt {2}} \frac{P_{\rm e}^{2} 
T_{\rm e}^{3/2}}{Emd_{\rm t}(T_{\rm e})}~.  % (5)
\end{equation}
From eqn. (5) it is clear that one cannot use a boundary condition that
sets $F_{\rm c}(T_{\rm e})$ to exactly zero at some base temperature, since 
this would lead to an infinite value of $Emd_{\rm t}(T_{\rm e})$.

As carried out in earlier work [see Jordan \& Brown (1981)], eqn. (5) can
be differentiated to give 
\begin{equation}
\frac{{\rm d}\log Emd_{\rm t}(T_{\rm e})}{{\rm d}\log T_{\rm e}} = 3/2 + 
                      \frac{2 {\rm d}\log P_{\rm e}}{{\rm d}\log T_{\rm e}}
    - \frac{{\rm d}\log (-F_{\rm c}(T_{\rm e}))}{{\rm d}\log T_{\rm e}}~.% (6)
\end{equation}

So far, no assumptions have been made about the energy balance. 
It is now assumed that the corona is heated by a flux of energy from
lower layers and that this energy is not dissipated until high in the corona, 
far above the first pressure-squared isothermal scale-height, $H$, over which 
the spatially averaged stellar emission lines are mainly formed. (The same 
situation is relevant to solar lines observed near Sun-centre.) In this case, 
the heated region can be studied only through solar observations above the 
limb. In the Sun, $T_{\rm e}$ rises slowly up to a height of 0.70 $R_{\odot}$,
far higher than $H \simeq 0.06 R_{\odot}$, while the non-thermal velocities 
(interpreted using $T_{\rm e}$, rather than the unknown ion temperatures 
$T_{\rm i}$), continue to increase (Landi, Feldman \& Doschek, 2006). 

Thus, below the heated region, it can be assumed that the divergence of the 
thermal conductive flux is balanced by the radiation losses. If any 
dissipation of the mechanical heating flux ($F_{\rm m}$) {\it were} present 
in the upper transition region and inner corona, this would add a term 
$-{\rm d}F_{\rm m}/{\rm d}h$ to the right-hand side (RHS) of eqn. (7) below. 
This would result in a steeper 
$Emd_{\rm t}(T_{\rm e})$ (Jordan 2000), since it adds a positive term to the 
RHS of eqn. (10) below. An enthalpy flux is not included in either our full
numerical solutions or the analytical approximations, because large enough 
systematic flows are not usually observed in the upper transition region and 
inner corona. Hence
\begin{equation}
\frac{{\rm d}F_{\rm c}(T_{\rm e})}{{\rm d}h} = 
           - \frac{{\rm d}F_{\rm r}(T_{\rm e})}{{\rm d}h}~, % (7)
\end{equation}
where the radiation losses are given by
\begin{equation}
\frac{{\rm d}F_{r}(T_{\rm e})}{{\rm d}h} = 
      \frac{0.86 P_{\rm e}^{2} P_{\rm rad}(T_{\rm e})}{T_{\rm e}^2}~,  % (8)
\end{equation}
and where $P_{\rm rad}(T_{\rm e})$ is the radiative powerloss function. \\
 Or, using eqn. (3) for ${\rm d}h/{\rm d}T_{\rm e}$, 
\begin{equation}
\frac{{\rm d}F_{\rm r}(T_{\rm e})}{{\rm d}T_{\rm e}} = \frac{\sqrt {2} 
       Emd_{\rm t}(T_{\rm e})P_{\rm rad}(T_{\rm e})}{T_{\rm e}}~. % (9) 
\end{equation}

Eqn. (6) then becomes
\begin{equation}
\frac{{\rm d}\log Emd_{\rm t}(T_{\rm e})}{{\rm d}\log T_{\rm e}} = 3/2 + 
\frac{2 {\rm d}\log P_{\rm e}}{{\rm d}\log T_{\rm e}}  - 
\frac{2 Emd_{\rm t}(T_{\rm e})^{2} P_{\rm rad}(T_{\rm e})}{0.86 \kappa 
P_{\rm e}^{2} T_{\rm e}^{3/2}}~. % (10)  
\end{equation}
Here we approximate $P_{\rm rad}(T_{\rm e})$ by $\alpha T_{\rm e}^{-1/2}$, 
where $\alpha$ is a constant (taken as $2.8\times 10^{-19}$~
erg~cm$^{3}$~s$^{-1}$~K$^{1/2}$), on the grounds that we are considering only 
collisionally excited lines where 
$2 \times 10^5$~K $\le T_{\rm e} \le 10^7$~K. At higher temperatures 
continuum processes cause an increase in $P_{\rm rad}(T_{\rm e})$.
 Alternatively, numerical values can be used, for example from CHIANTI (v6) 
(Dere et al. 2009). The form adopted here is useful in elucidating the 
physics since it results in the simplest analytical relations. 

The variation of the total pressure (including a non-thermal 
pressure term) with $T_{\rm e}$ is included in our numerical solutions using  
hydrostatic equilibrium. In the present paper, the non-thermal pressure is 
neglected, since in $\epsilon$ Eri, above $T_{\rm e} = 2 \times 10^5$~K, it 
does not exceed 0.08 of the gas pressure (Sim, 2002). Hydrostatic 
equilibrium should be a good approximation in the upper transition region and 
inner corona, since flows with velocities approaching the sound speed are 
not usually observed in these regions. 

The third term in eqn. (10) is given by hydrostatic equilibrium, and can be 
expressed as,
\begin{equation}
2 \frac{{\rm d} \log P_{\rm e}}{{\rm d} \log T_{\rm e}} = 
     -2 \frac{\sqrt {2} \mu m_{\rm H}  Emd_{\rm t}(T_{\rm e}) g_{*} T_{\rm e} 
R_{*}^{2}}{0.86 {\rm k_{B}} P_{\rm e}^{2} (R_{*} + h)^{2}}~,  % (11)  
\end{equation}
where $\mu$ is the mean molecular weight, taken to be 0.619, and $R_{*}$ and 
$g_{*}$ are the stellar radius and surface gravity, respectively. The 
variation in this term is small in the mid transition region, but becomes more 
important as $T_{\rm e}$ increases, since 
$Emd_{\rm t}(T_{\rm e}) T_{\rm e}/P_{\rm e}^2$ increases with $T_{\rm e}$.

Although we drop the variation of $P_{\rm e}$ with $T_{\rm e}$ in
general, we do use the integral of ${\rm d}P_{\rm e}/{\rm d}T_{\rm e}$ to
relate $P_{\rm o}$ and $P_{\rm c}$, using  
\begin{equation}
P_{\rm o}^{2} =  P_{\rm c}^2 + \frac{2 \sqrt {2} \mu m_{\rm H} g_{*}}{0.86 
{\rm k_{B}}} \int_{T_{\rm o}}^{T_{\rm c}} Emd_{\rm t}(T_{\rm e}) \frac{R_{*}^2}
{(R_{*} + h)^{2}}~{\rm d}T_{\rm e}~.  % (12)
\end{equation}
To find an analytical solution (see Section 2.2) the variation of the gravity
 with height has to be neglected. The full numerical solutions available for 
dwarf stars (e.g. Philippides 1996; Sim 2002; Ness \& Jordan 2008) show that
 this variation is not very large in G/K dwarfs. E.g., in the final
numerical model by Ness \& Jordan (2008), using a radial height of 
$\le 3000$~km at $\log T_{\rm e}$ = 5.30, gives $g(r_{\rm o})/g_{*} \ge 0.99$
 and $g(r_{\rm c})/g_{*} \ge 0.90$ at $\log T_{\rm c}$ = 6.53.      

From eqn. (10), ignoring the variation in pressure, a {\it minimum} in 
$Emd_{\rm t}(T_{\rm e})$ occurs at some $T_{\rm o}$, when
\begin{equation}
Em_{\rm t}(T_{\rm o}) = \sqrt{\frac{3 \times 0.86 \kappa}{4 \alpha}} P_{\rm o}
                         T_{\rm o}~. % (13)
\end{equation}
Thus at a given $T_{\rm o}$, $Em_{\rm t}(T_{\rm o}) \propto P_{\rm o}$. 

\subsection{Global constraints and resulting scalings}
At the top of the transition region/base of the corona, $F_{\rm c}(T_{\rm c})$ 
is the energy conducted back from the overlying heated corona.  
At $T_{\rm o}$ the conductive flux is $F_{\rm c}(T_{\rm o})$.
The easiest way to use the global constraint that the overall net conductive
flux is balanced by the total radiation losses is to use the approach
of Rosner, Tucker \& Vaiana (1978), but in the present work no explicit 
boundary conditions on the 
values of $F_{\rm c}(T_{\rm c})$ and $F_{\rm c}(T_{\rm o})$ have been imposed.

From eqn. (7) one can write
\begin{equation}
\int_{T_{\rm o}}^{T_{\rm c}} F_{\rm c}(T_{\rm e}) {\rm d}F_{\rm c}(T_{\rm e})
= - \int_{T_{\rm o}}^{T_{\rm c}} F_{\rm c}(T_{\rm e}) \frac{{\rm d}F_{\rm r}}
{{\rm d}h} \frac{{\rm d}h}{{\rm d}T_{\rm e}} {\rm d}T_{\rm e}~.   % (14)
\end{equation} 
Hence, using eqns (4) and (8), 
\begin{equation}
\frac{1}{2} (F_{\rm c}(T_{\rm c})^{2} - F_{\rm c}(T_{\rm o})^{2}) =
0.86 \alpha \kappa \int_{T_{\rm o}}^{T_{\rm c}} P_{\rm e}^{2} {\rm d}T_{\rm e}
~.        % (15)
\end{equation}  
Then, on the left-hand side of eqn. (15), eqn. (3) can be used to express
${\rm d}T_{\rm e}/{\rm d}h$ in terms of $Emd_{\rm t}(T_{\rm e})$. On the RHS 
of eqn. (15) the pressure term is taken to be 
constant at $P_{\rm o}^2$. This is close to the mean value of $P_{\rm e}^{2}$ 
in the full numerical solutions. The pressure variation between $T_{\rm o}$ 
and $T_{\rm c}$ {\it is} included in the conductive flux terms. The result is 
\begin{equation}
\frac{0.86 \kappa}{4} (\frac{T_{\rm c}^{3} P_{\rm c}^{4}}{Em_{\rm t}
(T_{\rm c})^{2}} - \frac{T_{\rm o}^{3} P_{\rm o}^{4}}{Em_{\rm t}(T_{\rm o})
^{2}}) =  \alpha  P_{\rm o}^{2} (T_{\rm c} - T_{\rm o})~.        % (16)  
\end{equation} 

Substituting for $Em_{\rm t}(T_{\rm o})$ 
from eqn. (13), eqn. (16) can be rearranged to give 
\begin{equation}
Em_{\rm t}(T_{\rm c}) = \sqrt{\frac{0.86 \kappa}{4 \alpha }}\frac{T_{\rm c}
P_{\rm c}^{2}}{P_{\rm o}}\frac{1}{(1 - \frac{2T_{\rm o}}{3T_{\rm c}})^{1/2}}~,
  % (17)  
\end{equation}
where $T_{\rm c}$ can be replaced by a general $T_{\rm e}$ ($> T_{\rm o}$), 
and similarly, $P_{\rm c}$ can be replaced by a general $P_{\rm e}$, to give a
 general equation for $Emd_{\rm t}(T_{\rm e})$. 

$Em_{\rm t}(T_{\rm c})$ can also be found by making the approximation that 
the coronal emission is formed mainly over the first pressure-squared 
isothermal scale-height.
This is justified by using the equation of hydrostatic equilibrium to show
that $P_{\rm e}^{2}$ decreases exponentially with height according to 
$P_{\rm c}^{2} \exp[-(h - h_{\rm c})/H]$, where $h_{\rm c}$ is the height at 
the base of the corona and $H = {\rm k_{B}} T_{\rm c}/2 \mu {\rm m_H} g_{*}$. 
Solar observations by Gibson et al. (1999) confirm the
hydrostatic decrease in $N_{\rm e}^2$ under the near-isothermal conditions in
the low corona. In this case, 
     
\begin{equation}
Em_{\rm t}(T_{\rm c}) = \frac{0.86 {\rm k_{B}}}{2 \mu m_{\rm H}} 
\frac{P_{\rm c}^{2}}{g_{*} T_{\rm c}}.   % (18)
\end{equation} 
Combining eqns (17) and (18) allows $P_{\rm c}$ to be eliminated, leading to
an explicit expression for $P_{\rm o}$,
\begin{equation}
P_{\rm o} = \sqrt{\frac{\kappa}{0.86 \alpha}} \frac{\mu m_{\rm H}}{{\rm k_{B}}}
\frac{g_{*} T_{\rm c}^{2}}{(1 - \frac{2T_{\rm o}}{3T_{\rm c}})^{1/2}}~. % (19)
\end{equation}
Thus $P_{\rm o}$ scales as $g_{*} T_{\rm c}^{2}$, with
only a weak dependence on $T_{\rm o}/T_{\rm c}$, and provided $g_{*}$ is known,
can be found from the values of $T_{\rm o}$ and $T_{\rm c}$ that match the
observed behaviour of $Emd_{\rm app}(T_{\rm e})$.

Once $P_{\rm o}$ is known, $Em_{\rm t}(T_{\rm o})$ is known from eqn. (13) and
 can also be expressed in terms of $g_{*}$, $T_{\rm c}$ and $T_{\rm o}$, i.e. 
\begin{equation}
Em_{\rm t}(T_{\rm o}) = \sqrt{3} \frac{\kappa \mu m_{\rm H}}{2 \alpha 
{\rm k_{B}}} \frac{g_{*} T_{\rm o} T_{\rm c}^{2}}
{(1 - \frac{2T_{\rm o}}{3 T_{\rm c}})^{1/2}}~. %  (20)
\end{equation}   

The ratio of $Em_{\rm t}(T_{\rm c})$ to $Em_{\rm t}(T_{\rm o})$ is given by 
\begin{equation}
\frac{Em_{\rm t}(T_{\rm c})}{Em_{\rm t}(T_{\rm o})} = 
\frac{T_{\rm c}}{T_{\rm o}}\frac{P_{\rm c}^{2}}{P_{\rm o}^{2}} \frac{1}
{\sqrt{3} (1 - \frac{2 T_{\rm o}}{3 T_{\rm c}})^{1/2}}~. % (21)
\end{equation}
To separate $Em_{\rm t}(T_{\rm c})$ and $P_{\rm c}^2$ requires eqn. (12), 
which can be written as 
\begin{equation}
\frac{P_{\rm o}^{2}}{P_{\rm c}^{2}} = 1 + \frac{\sqrt{2} \int_{T_{\rm o}}^
{T_{\rm c}} Emd_{\rm t}(T_{\rm e})~{\rm d}T_{\rm e}}{Em_{\rm t}(T_{\rm c}) 
T_{\rm c}}~.     % (22) 
\end{equation}
The detailed numerical models show that $P_{\rm o}^{2}/P_{\rm c}^{2}$ varies 
slowly when $T_{\rm c}$ is varied, because the last term in eqn. (22) is 
almost constant. The general form of $Emd_{\rm t}(T_{\rm e})$, when 
$T_{\rm c}$ in eqn. (17) is replaced by $T_{\rm e}$, can now be applied. This
 time, guided by the full calculations, we take the variable $P_{\rm e}^2$ as
 $P_{\rm c}^2$ and remove it from the integral. The integral of 
$T_{\rm e}/[1- (2/3)(T_{\rm o}/T_{\rm e})]^{1/2}$ has a standard solution 
(see e.g. Jeffrey 1995), which reduces to 
\begin{equation}    
 \frac{ 8 T_{\rm o}^{2}}{9} \int \frac{1}{(1 - Y^{2})^{3}} {\rm d}Y = 
      \frac{T_{\rm o}^2}{9}\frac{Y(5-3Y{^2})}{(1-Y^{2})^{2}} 
- \frac{T_{\rm o}^{2}}{6} \ln{\left[\frac{|Y-1|}{|Y+1|}\right]},    % (23) 
\end{equation}
where $Y^{2} = 1 - (2T_{\rm o}/3T_{\rm e})$. The first term on the RHS
of eqn. (23), evaluated
at $T_{\rm c}$, dominates and reproduces the full solution to within 
0.1~per cent. Thus, when the constants in the general expressions for 
$Emd_{\rm t}(T_{\rm e})$ and $Em_{\rm t}(T_{\rm c})$ are included, the last 
term in eqn. (22) becomes
\begin{equation}
\frac{\sqrt{2}\int_{T_{\rm o}}^{T_{\rm c}} Emd_{\rm t}(T_{\rm e}) {\rm d}
T_{\rm e}}{Em_{\rm t}(T_{\rm c}) T_{\rm c}} = \frac{1}{\sqrt{2}}
(1-\frac{2T_{\rm o}}{3T_{\rm c}})(1 + \frac{T_{\rm o}}{T_{\rm c}})~. % (24) 
\end{equation}
Hence $P_{\rm o}^{2}/P_{\rm c}^{2}$ can be found from eqns (22) and (24), and 
$P_{\rm c}^{2}$ can then be found, using 
$P_{\rm o}$ from eqn. (19). 

Using eqn. (18) $Em_{\rm t}(T_{\rm c})$ can then 
be expressed as  
\begin{equation}
Em_{\rm t}(T_{\rm c}) = \frac{\mu m_{\rm H}\kappa}{2 {\rm k_{B}}\alpha} 
\frac{g_{*} T_{\rm c}^{3}}{(1 - \frac{2T_{\rm o}}{3 T_{\rm c}})[1 + \frac{1}
{\sqrt{2}}(1-\frac{2 T_{\rm o}}{3 T_{\rm c}})(1 + \frac{T_{\rm o}}
{T_{\rm c}})]}~. % (25) 
\end{equation}
This gives the scaling with $g_{*} T_{\rm c}^{3}$, expected from dimensional
arguments, and also the absolute value, which arises from the energy balance
adopted. The terms in $T_{\rm o}/T_{\rm c}$ vary by only 1~per cent over the 
range of $T_{\rm c}$ in the models discussed below. Thus, for a given choice 
of $\kappa$ and $\alpha$, only $g_{*}$, 
$T_{\rm c}$ and $T_{\rm o}$ are required to evaluate $Em_{\rm t}(T_{\rm c})$.

We now make comparisons with results from Hearn's (1975, 1977) formulation
 for minimum energy loss (mel) coronae. Hearn applied ${\rm d}F_{\rm c} = -
{\rm d}F_{\rm r}$ to a corona, but took a partial differentiation with 
respect to the coronal temperature, at constant coronal pressure. He assumed 
that $T_{\rm o}$ and $F_{\rm c}(T_{\rm o})$ can be neglected.

Making the same assumptions, by dropping the second term in eqn. (15) and 
using eqn. (18), it can be seen that
\begin{equation}
P_{\rm o}({\rm mel}) = \sqrt{\frac{\kappa}{0.86 \alpha}}
                \frac{\mu m_{\rm H}}{{\rm k_{B}}} g_{*} T_{\rm c}^2.    % (26)
\end{equation}       

However, in our approach, if $F_{\rm c}(T_{\rm o})$ tends to zero, then 
$Em_{\rm t}(T_{\rm o})$ tends to infinity. Thus eqn. (16), together with 
eqn. (18) can be used to define a critical (maximum) value of 
$P_{\rm o}$. This pressure cannot be reached when there is a minimum in 
$Emd_{\rm t}(T_{\rm e})$ at $T_{\rm o}$. $P_{\rm o}({\rm crit})$ is given by
\begin{equation}
P_{\rm o}({\rm crit}) = \sqrt{\frac{\kappa}{0.86 \alpha}} \frac{\mu m_{\rm H}}
{{\rm k_{B}}} \frac{g_{*} T_{\rm c}^{2}}{(1 - \frac{T_{\rm o}}{T_{\rm c}})
^{1/2}}~. % 27
\end{equation}   
This approaches the value of $P_{\rm o}({\rm mel})$ when $T_{\rm o}$ is much 
smaller than $T_{\rm c}$. If observations of density/pressure sensitive lines 
show clearly that $P_{\rm o}$ {\it exceeds} $P_{\rm o}({\rm crit})$, then 
models including a fractional emitting area, or heating, additional to that 
provided by thermal conduction, must be considered.    

Although the difference between our predicted values of $P_{\rm o}$ and 
$P_{\rm o}(\rm mel)$ are not large, the advantage of our solutions is that 
they allow for the difference between $P_{\rm o}$ and $P_{\rm c}$ in 
hydrostatic equilibrium and predict the values of $Em_{\rm t}(T_{\rm o})$, as 
well as $Em_{\rm t}(T_{\rm c})$. Hearn (1975, 1977) assumed a constant pressure
with a value defined at the `base of the corona'. Applying the pressure from 
eqn. (26) to find $Em_{\rm t}(T_{\rm c})$ from eqn. (18) gives larger values
 than found from our analytical solutions.    

\section{Results and comparisons with full models}
In optimizing our full solution for the $Emd_{\rm t}(T_{\rm e})$ and 
$Emd_{\rm app}(T_{\rm e})$ of 
$\epsilon$~Eri (Ness \& Jordan 2008), we ran five models with different 
values of $T_c$. The results for the full models are those for which a 
minimum in $Emd_{\rm app}(T_{\rm e})$ is just possible at the chosen value of 
$T_{\rm o} = 2 \times 10^5$~K. Such solutions are found by gradually 
increasing the value of $Em_{\rm app}(T_{\rm c})$. As stressed earlier, the
full numerical models are in hydrostatic equilibrium, including a 
non-thermal pressure term based on observed line widths, and adopt a 
spherically
symmetric atmosphere. Although we do not expect exact agreement between the
analytical and the full models, it is of interest to examine the size of the
differences between them, since the analytical models can be useful in
the process of homing in on the optimum solution for a given star.

Table 1 gives the values of the parameters discussed in the previous section,
with those from the above full numerical models given in the lines labelled 
`n' and those from the analytical predictions
in the lines labelled `a'. The order of the parameters listed reflects the 
order in which the calculations can be made, i.e., $P_{\rm o}$ from eqn. (19), 
$Em_{\rm t}(T_{\rm o})$ from eqn. (13),  $(P_{\rm o}/P_{\rm c})^2$ from eqns 
(22) and (24), $P_{\rm c}$ from eqns (19), (22) and (24), and 
$Em_{\rm t}(T_{\rm c})$ from using  $P_{\rm c}^2$ in eqns (17) or (18).

%Table 1
\begin{table}
\caption{Comparison of parameters from the numerical models for $\epsilon$~Eri
(lines labelled `n') and from the analytical equations (lines labelled `a'). 
The quantities involved, including their units are: 
$\log [T_{\rm e}({\rm K})]$, $\log [P_{\rm e}$(cm$^{-3}$~K)] and 
$\log [Em_{\rm t}(T_{\rm e})$(cm$^{-5}$)]; $\log T_{\rm o}$ is fixed at 
5.30.}  
\begin{tabular}{lcccccc}
\hline \\
$\log T_{\rm c}$  &      &  6.50  &  6.53  &  6.55  &  6.60  &  6.65  \\
\hline \\
$\log P_{\rm o}$ & (n)  & 15.91 & 15.97 & 16.01 & 16.10 & 16.20    \\
$\log P_{\rm o}$ & (a)  & 15.86 & 15.92 & 15.96 & 16.06 & 16.16\\
$\log Em_{\rm t}(T_{\rm o})$ & (n) & 27.50 & 27.56 & 27.60 & 27.69 & 27.79 \\ 
$\log Em_{\rm t}(T_{\rm o})$ & (a) & 27.37 & 27.43 & 27.47 & 27.56 & 27.66 \\
$2\log (P_{\rm o}/P_{\rm c})$ & (n) & 0.254 & 0.254 & 0.253 & 0.252 & 0.251 \\
$2\log (P_{\rm o}/P_{\rm c})$ & (a) & 0.236 & 0.235 & 0.235 & 0.235 & 0.235 \\
$\log P_{\rm c}$ & (n)  & 15.79 & 15.84 & 15.88 & 15.98 & 16.07    \\
$\log P_{\rm c}$ & (a)  & 15.75 & 15.81 & 15.85 & 15.95 & 16.04    \\
$\log Em_{\rm t}(T_{\rm c})$ & (n) & 28.22 & 28.31 & 28.36 & 28.51 & 28.66  \\
$\log Em_{\rm t}(T_{\rm c})$ & (a) & 28.10 & 28.19 & 28.25 & 28.40 & 28.55  \\
\hline   \\
\end{tabular}
\end{table}

The full numerical models give the same scaling laws for $P_{\rm o}$ and 
$P_{\rm c}$ as found from the analytical approach, but with multiplication
 factors of about 1.1. Similarly, the full numerical emission measures 
[$Em_{\rm t}(T_{\rm o})$ and $Em_{\rm t}(T_{\rm c})$] follow the same scaling 
laws as found from the analytical approach, but are systematically larger by 
mean factors of about 1.3.  The differences arise from the approximations to 
the electron pressure used in the analytical equations, including the neglect 
of the non-thermal pressure term, and to the different geometries adopted. The
variation of $g$ with the radial distance $r$ is also included in the 
numerical solutions. 

Table 2 gives the combinations of parameters that appear in eqn. (21)    
and in the fourth term of eqn. (10).

% Table 2
\begin{table}
\caption{Values of $\log[Em_{\rm t}(T_{\rm c})/Em_{\rm t}(T_{\rm o})]$ and 
$Em_{\rm t}(T_{\rm c})/P_{\rm c} T_{\rm c}$ calculated from the full numerical
 models 
for $\epsilon$~Eri (lines labelled `n') and the analytical solutions (lines 
labelled `a'). Units as in Table 1.}
\begin{tabular}{lcccccc}
\hline \\
$\log T_{\rm c}$ &                      & 6.50 & 6.53 & 6.55 & 6.60 & 6.65   \\
\hline \\
$\log [Em_{\rm t}(T_{\rm c})/Em_{\rm t}(T_{\rm o})]$ & (n) & 0.72 & 0.74 & 
0.76 & 0.82 & 0.87  \\
$\log [Em_{\rm t}(T_{\rm c})/Em_{\rm t}(T_{\rm o})]$ & (a) & 0.73 & 0.76 & 
0.78 & 0.83 & 0.88 \\
$\log[Em_{\rm t}(T_{\rm c})/P_{\rm c} T_{\rm c}]$ & (n) & 5.93 & 5.93 & 5.93 &
 5.93 & 5.94  \\
$\log[Em_{\rm t}(T_{\rm c})/P_{\rm c} T_{\rm c}]$ & (a) & 5.85 & 5.85 & 5.85 &
 5.85 & 5.85   \\
\hline \\
\end{tabular}
\end{table}

It can be seen that the ratio of the emission measures at $T_{\rm c}$ and 
$T_{\rm o}$ given by eqn. (21) agrees better with the results from the full 
models than do the absolute values. The numerical models show that the ratio
 $Em_{\rm t}(T_{\rm c})/P_{\rm c} T_{\rm c}$ is almost constant, as expected 
from the scaling laws, 
$Em_{\rm t}(T_{\rm c}) \propto g(r_{\rm c}) T_{\rm c}^{3}$ and 
$P_{\rm c} \propto g(r_{\rm c}) T_{\rm c}^{2}$. 

In the numerical solutions for $\epsilon$~Eri, the origins of the values of 
the constants of proportionality in the empirical scaling laws are hard to 
pin down. The analytical expressions give similar scalings, but now the 
actual value of the constant of proportionality can be clearly tracked back 
to the global energy balance equation assumed.
  
\section{Example application}
Here we apply the analytical expressions to Procyon (HD 61421, F5 IV-V) to 
illustrate what can be learnt before detailed modelling is carried out. 
Earlier work on Procyon has shown that it is difficult to reconcile different 
measurements of $P_{\rm e}$ without invoking limited areas of emission
or the presence of active region loops (Schmitt et al. 1985, 1996; Jordan et
 al. 1986). Because of the lower gravity and larger line widths in
 Procyon, compared with those of cool dwarf stars, the simple methods used 
in Section 2 are expected to be less accurate than for $\epsilon$~Eri. 
For a quiescent corona, the nature of the heating flux is
not relevant in our energy balance model, so either an MHD wave flux or an
acoustic wave flux, as suggested by Mullen \& Cheng (1994), is possible.

Emission line fluxes are available for Procyon from a number of the instruments
mentioned in Section 1. Jordan et al. (1986) analyzed spectra obtained with 
the {\it IUE} and the {\it Einstein Observatory} and derived emission 
measures, but not a mean
emission measure distribution. Early observations with {\it Copernicus} were 
used to constrain the line emission measure around $3 \times 10^5$~K. They
 found that the ratio of the coronal emission measure to that of 
the lower transition region was significantly smaller than in main-sequence 
dwarf stars and that the mean coronal temperature was around 
$1.5 \times 10^6$~K. This was in
broad agreement with earlier work by Schmitt et al. (1985), who used data
from the {\it Einstein Observatory}. In particular, even with the uncertainty
 in the line fluxes from {\it Copernicus}, it appeared that $T_{\rm o} \ge 
3 \times 10^5$~K, rather than $2 \times 10^5$~K in the main-sequence stars. 
Drake et al. (1995) included data obtained with the {\it EUVE}, plus some 
adjusted data from {\it Copernicus}, to produce the emission measure
distribution above $T_e \simeq 1.6 \times 10^5$~K, while Sanz-Forcada, 
Brickhouse \& Dupree (2003) combined the observations from the {\it IUE} and 
the {\it EUVE} to improve the overall $Emd_{\rm app}(T_{\rm e})$. Although 
there were 
differences in detail, owing to the abundances and atomic data adopted, the 
$Emd_{\rm app}(T_{\rm e})$ found by Sanz-Forcada et al. (2003) showed a 
similar form to 
that indicated in Jordan et al. (1986), but with $T_{\rm o} \simeq 4 \times 
10^5$~K. Sanz-Forcada et al. (2004), improved the higher temperature part of 
the EMD using spectra obtained with the LETGS on the {\it Chandra} satellite. 
They found 
$T_{\rm o} \simeq 5 \times 10^5$~K and $T_{\rm c} \simeq 2 \times 10^6$~K. 
Raassen et al. (2002) also analyzed spectra from the LETGS and the 
{\it XMM-Newton} satellite, but made a global 3-T fit using the SPEX computer 
package, rather than an individual line-based approach. Wood et al. (1996) 
analyzed the line widths and redshifts measured from spectra obtained with the
GHRS instrument on the {\it HST}. The line widths were found to be larger  
than in the main-sequence stars, such as $\epsilon$~Eri (Sim \& Jordan 2003b).
 In summary, in Procyon, $\log [T_{\rm o}({\rm K})]$ lies 
between 5.5 and 5.7 and $\log [T_{\rm c}({\rm K})]$ lies between 6.2 and 6.3.  
The equation of hydrostatic equilibrium should strictly include the effects 
of any non-thermal pressure associated with the larger line widths observed.  

We adopt the following stellar properties: a distance of 3.53~pc (Girard et 
al. 2000), an angular diameter of 5.51~mas (Mozurkewich et al. 1991) and
 hence $R_{*} = 2.09 R_{\odot}$; a mass of $M_{*} = 1.5 M_{\odot}$ (Girard et
 al. 2000) and hence $\log [g_{*}$(cm~s$^{-2})] = 3.98$.  

The results of applying the analytical expressions given in Section 2 are 
summarized in Table 3, for $\log [T_{\rm o}({\rm K})]$ = 5.5 and 5.7, and 
$\log [T_{\rm c}({\rm K})]$ = 6.2, 6.25 and 6.3.  

%Table 3
\begin{table}
\caption{Predicted analytical values of $P_{\rm o}$, $P_{\rm c}$, 
$Em_{\rm t}(T_{\rm o})$
and $Em_{\rm t}(T_{\rm c})$ for Procyon, together with combined parameters.
 Results using $\log T_{\rm o} = 5.5$ are given in the upper part of the 
table, and those using $\log T_{\rm o} = 5.7$ in the lower part. Units as in
 Table 1.}  
\begin{tabular}{lccc}
\hline \\
$\log T_{\rm c}$              &   6.20   &  6.25   &    6.30    \\
\hline \\
$\log T_{\rm o}$ = 5.5        &          &         &              \\ 
$\log P_{\rm o}$              &  14.62   & 14.71   &   14.81     \\
$\log P_{\rm c}$              &  14.50   & 14.59   &   14.69     \\
$\log Em_{\rm t}(T_{\rm o})$  &  26.32   & 26.41   &   26.51     \\
$\log Em_{\rm t}(T_{\rm c})$  &  26.57   & 26.71   &   26.86     \\
$\log [Em_{\rm t}(T_{\rm c})/P_{\rm c} T_{\rm c}]$ & 5.87 & 5.87 & 5.87 \\
                              &          &         &              \\
$\log T_{\rm o}$ = 5.7        &          &         &              \\
$\log P_{\rm o}$              &  14.64   &  14.73  &   14.82     \\
$\log P_{\rm c}$              &  14.52   &  14.61  &   14.70     \\
$\log Em_{\rm t}(T_{\rm o})$  &  26.54   &  26.63  &   26.73       \\
$\log Em_{\rm t}(T_{\rm c})$  &  26.61   &  26.75  &   26.89      \\
$\log [Em_{\rm t}(T_{\rm c})/P_{\rm c} T_{\rm c}]$  & 5.89 &5.89 & 5.89 \\
\hline \\
\end{tabular}
\end{table}

As yet there are no completely satisfactory numerical models of the outer 
atmosphere of Procyon. Philppides (1996) used observations from 
{\it ROSAT} ({\it R\"Oentgen SATellit})  to find 
$Em_{\rm app}(T_{\rm c})$ and $T_{\rm c}$, and made models in a spherically 
symmetric geometry, including a non-thermal pressure term. She noted that the 
latter term causes $P_{\rm e}$ to {\it increase} with $T_{\rm e}$ within the 
transition region, before decreasing again by $T_{\rm c}$. The later 
observations with {\it EUVE} (Drake et al. 1995; Schmitt et al. 1996) and 
both the LETGS and {\it EUVE} (Sanz-Forcada et al. 2004) showed that, as 
expected, the earlier 1-temperature and 2-temperature fits to {\it ROSAT}
spectra overestimated the coronal emission measure. Sim (1998, unpublished 
MPhys project; 2002) made a model of the chromosphere and lower transition 
region using line fluxes and widths measured by Wood et al. (1996), which 
supersede the fluxes from {\it Copernicus} by Jordan et al. (1986), and 
included the radiative transfer in lines formed up to 
$\simeq 2 \times 10^4$~K. He adopted a plane parallel atmosphere up to 
$3 \times 10^5$~K and a spherically symmetric atmosphere at higher 
temperatures, including the non-thermal pressure term throughout the 
atmosphere. However, he interpolated $Emd_{\rm app}(T_{\rm e})$ between 
$\simeq 3 \times 10^5$~K and a coronal temperature of $2 \times 10^6$~K, 
rather than making an energy balance model. 

Here we use the work by Sanz-Forcada et al. (2004) to give information on
$Emd_{\rm app}(T_{\rm e})$, but adopt an iron abundance of 
$\log N_{\rm Fe}$ = 7.51, rather than their value of 7.32 (on the scale where
 $\log N_{\rm H}$ is 12.0). 
Their absolute scale has been changed to match our definition of 
$Em_{\rm t}(0.3)$ given by eqn. (1). Their maximum value of 
the apparent volume emission measure at $\log [T_{\rm c}({\rm K})] = 6.30$ 
has been divided by 
$4 \pi R_{*}^2$ to find $Em_{\rm app}(T_{\rm c})$. Because of this conversion,
in a spherically symmetric geometry the apparent and true emission measures 
are related by 
\begin{equation}
 Em_{\rm app}(T_{\rm e}) = f(r)G(r) Em_{\rm t}(T_{\rm e})~,     % (28)
\end{equation}
where $f(r) = (r/R_{*})^2$ and $G(r) = 0.5(1 + \sqrt{[1 - (1/f(r))]})$ is the
fraction of emitted photons not intercepted by the star. Thus in an extended
atmosphere the product $f(r)G(r)$ can be greater than 1, so that 
$Em_{\rm app}(T_{\rm c})$ can be larger than $Em_{\rm t}(T_{\rm c})$.   

Table 4 gives the observed values of 
$Em_{\rm app}(T_{\rm o})$ and $Em_{\rm app}(T_{\rm c})$ from Sim (1998) and 
Sanz-Forcada et al. (2004), respectively. Values of $\log P_{\rm o}$ and 
$\log P_{\rm c}$ from Sim's (2002) numerical models are also given.

% Table 4
\begin{table}  
\caption{Values of $P_{\rm o}$, $P_{\rm c}$,  $Em_{\rm app}(T_{\rm o})$ and 
$Em_{\rm app}(T_{\rm c})$, from numerical models and observations. 
Units as in Table 1.}
\begin{tabular}{lccc}
\hline \\
$\log T_{\rm c}$                &   6.30        &           &\\
\hline \\
$\log T_{\rm o}$ = 5.5          &               &           &\\
$\log P_{\rm o}$                &  14.81$^a$    &           &\\
$\log P_{\rm c}$                &  14.73$^a$    &           & \\
$\log Em_{\rm app}(T_{\rm o})$  &  26.27$^b$    &           &\\
$\log Em_{\rm app}(T_{\rm c})$  &  26.84$^c$    &           & \\
                                &  27.16$^d$    &           & \\
\hline \\
\end{tabular}

$^a$ From numerical model by Sim (2002). \\
$^b$ From observations fitted by Sim (1998). \\
$^c$ Starting value adopted by Sim (1998). \\
$^d$ From observations fitted by Sanz-Forcada et al. (2004), scaled as 
described in the text.
\end{table}
 
We can now compare the predicted values of $Em_{\rm t}(T_{\rm o})$ and 
$Em_{\rm t}(T_{\rm c})$, given in Table 3 for $\log [T_{\rm c}({\rm K})]$ = 
6.30, with the values of $Em_{\rm app}(T_{\rm o})$ and 
$Em_{\rm app}(T_{\rm c})$ given in Table 4. The value of 
$\log Em_{\rm app}(T_{\rm o})$, when $\log [T_{\rm o}({\rm K})] = 5.50$, 
found by Sim (1998) is smaller than the value of 
$\log Em_{\rm t}(T_{\rm o})$. In a plane parallel atmosphere these are 
expected to differ by about a factor of two, which is indeed the case. Bearing
 in mind the comparisons between the analytical and numerical results for 
$\epsilon$~Eri, this suggests that the emission at this height is {\it not} 
from a highly restricted 
area of the atmosphere. The value of $Em_{\rm t}(T_{\rm c})$ is similar to, or
less than the value of $Em_{\rm app}(T_{\rm c})$. Using the pressure-squared
isothermal scale height ($H$) in the corona, plus $R_*$, to estimate $r$ at 
$T_{\rm c}$ leads to 
$f(r)G(r) \simeq 0.85$. Given the uncertainties in, and differences between 
the atomic data used by the above authors, the disagreement between the 
predicted and observed values is not large and is in the
direction found from the numerical and analytical results for $\epsilon$~Eri. 
However, Procyon is a case where full numerical models are required to
make more detailed comparisons between observed and predicted results. 

It is very difficult to establish values of $N_{\rm e}$ in Procyon.
Schmitt et al. (1996) made a careful analysis of density sensitive lines of 
Fe\,{\sc x} to {\sc xiv} observed with the {\it EUVE} and concluded that the 
average value of $N_{\rm e}$ at coronal temperatures lies between 
$10^9$ and $10^{10}$~cm$^{-3}$, with a value of $3 \times 10^9$~cm$^{-3}$
 being adopted. However, for some ions the results depended on weak lines, 
and a range of $N_{\rm e}$ was found from different pairs of lines within a 
given ion. As Schmitt et al. (1996) point out, the uncertainties in 
$N_{\rm e}$ mask any systematic variation of $N_{\rm e}$ or 
$P_{\rm e}$ with the stage of ionization. At around 
$\log [T_{\rm c}({\rm K})] = 6.20 - 6.30$, their smallest value of 
$\log [P_{\rm c}$(cm$^{-3}$~K)] = 15.2 is significantly larger than those 
given in Table 3. 

Ness et al. (2001) measured the ratio of the fluxes in the forbidden line
(1s$^2$~$^1$S$_{0}$ - 1s2s~$^3$S$_{1}$) and intersystem (plus quadrupole) line 
(1s$^2$~$^1$S$_{0}$ - 1s2p~$^3$P$_{1,2}$) in the He\,{\sc i}-like ions 
C\,{\sc v}, N\,{\sc vi} and O\,{\sc vii}, using spectra obtained with the 
LETGS. We have found revised flux ratios by also including further spectra 
available from more recent observations with the LETGS. We have also updated 
the values of (and limits on) $N_{\rm e}$ by using CHIANTI (v6) (Dere et al. 
2009). In the Sun, only the flux ratio in C\,{\sc v} is affected by the 
photospheric/chromospheric radiation field at the wavelength of the 
$^3$S$_1$ - $^3$P transitions (Gabriel \& Jordan 1969). In Procyon, because 
of the higher photospheric temperature, Ness et al. (2001) found that this 
photoexcitation is also significant in N\,{\sc vi}. As a result, $N_{\rm e}$ 
is not constrained by the observed flux ratio in C\,{\sc v}. In N\,{\sc vi}, 
the combined uncertainties in the radiation field and the measured flux ratio 
are too large to yield a definitive value of $\log [N_{\rm e}$(cm$^{-3}$)], 
which can lie between $\le 8.0$ and 10.0.
 O\,{\sc vii} provides the best diagnostic, since the effects of the 
radiation field are small and make little difference to the value of 
$\log [N_{\rm e}$(cm$^{-3}$)] = 9.13 obtained (when these effects are 
included). But the uncertainty of about $\pm$ 10 per cent in the measured 
flux ratio of 3.60 includes the value of 3.75 at 
$\log [N_{\rm e}$(cm$^{-3}$)] = 8.0. The observed ratio yields 
$\log [P_{\rm e}$(cm$^{-3}$~K)] = 15.43, with an upper limit of 15.99 and a 
lower limit of $\le 14.3$. 

Liang, Zhao \& Shi (2006) investigated density-sensitive X-ray lines of 
Si\,{\sc x}, using LETGS fluxes from Raassen et al. (2002). From the strongest
 lines at 50.524 and 50.691~\AA\ they found $\log [N_{\rm e}$(cm$^{-3}$)] 
$\simeq 8.41 - 8.45$ at $T_{\rm e} = 1.26 \times 10^6$~K, and that
 calculations by other authors gave only slightly higher values.
The resulting values of $\log [P_{\rm e}$(cm$^{-3}$K)] were in the range 
14.5 -- 14.7. These are similar to the value found from Sim's (2002) numerical
 model and to the value predicted by the analytical scaling laws.
The Si\,{\sc x} lines are expected to be formed at the same $T_{\rm e}$ as 
those of N\,{\sc vi}. 
However, in Si\,{\sc x}, the high density limit for the flux ratio occurs
at a lower value of $N_{\rm e}$ than in N\,{\sc vi}. If the emitting regions
contained both a quiescent corona and active regions (of smaller area), then
the higher density regions could contribute relatively less to the electron 
density measured from Si\,{\sc x}. Thus one cannot rule out the possibility
that N\,{\sc vi} could detect a higher value of $N_{\rm e}$. 

Jordan et al. (1986) used simple line opacity arguments to deduce that at
$T_{\rm e} = 2 \times 10^5$~K, $\log [P_{\rm e}$(cm$^{-3}$~K)]$\le 14.4$. 
This is a factor of
 two lower than the value of 14.76 in the model by Sim (2002). Wood et 
al. (1996) made a more sophisticated estimate of line opacity effects and
deduced that the profile of the line of Si\,{\sc iii} at 1206~\AA\ implied a
pressure of $\log [P_{\rm e}$(cm$^{-3}$~K)]$ \le 14.8$ at 
$\log [T_{\rm e}$(K)] = 4.70, in 
reasonable agreement with the value of 14.68 in Sim's (2002) model. 
However, the intersystem lines of O\,{\sc iv} analyzed by Wood et al. (1996) 
led to $\log [P_{\rm e}$(cm$^{-3}$~K)]$\simeq 15.0$ (or $\le 15.5$, when 
possible errors  
were considered). But as they pointed out, the O\,{\sc iv} flux ratios were 
not far from those expected in the low-density limit. 
Analyses of spectra observed recently by Ayres\footnote{archive.stsci.edu}
 with the STIS should give at least improved limits on values of $P_{\rm e}$.  
The values of $\log P_{\rm o}$ from the methods used here are 
given in Table 3 and agree well with Sim's (2002) model value of 14.81.      

Overall, the higher coronal values of $P_{\rm e}$ found by Schmitt et 
al. (1996), and from our present analysis of O{\sc vii}, could be reconciled 
with the 
values from the analytical predictions and the numerical models if higher 
pressure active region loop structures were present, as well as a quiescent 
corona (see also Schmitt et al. 1985, Jordan et al. 1986). Such active regions
 could also contribute to $Emd_{\rm app}$ near its apparent maximum value. 
However, it seems very unlikely that the quiescent atmosphere has pressures
that significantly exceed those given in Tables 3 and 4. Whether or not the
upper part of the quiescent atmosphere is heated by acoustic or MHD waves is 
still an open question, but in the time-averaged acoustic heating model by 
Mullan \& Cheng (1994), the maximum coronal temperature is only 
6.5 $\times 10^5$~K. If this temperature is adopted for $T_c$, the predicted
value of $Em_{\rm t}(T_{\rm o})$ is about a factor of 5 smaller than the
observed value given in Table 4, so any acoustic heating must lead either 
to a hotter corona, or be limited to the region below $T_{\rm o}$. 

By applying the analytical solutions we have found that the value of 
$Em_{\rm t}(T_{\rm o})$ is acceptable without invoking a limited area of 
emission and that current acoustic heating models are not
entirely satisfactory. We have also predicted the values of $P_{\rm e}$ in the
quiescent transition region and corona. While the former is acceptable, 
higher pressures cannot be ruled out above about $T_{\rm e} \ge 10^{6}$~K. 
Procyon is clearly a star for which full numerical models are required.
These will be carried out following the methods used by Ness \& Jordan (2008)
in studies of $\epsilon$~Eri, and will include recent atomic data
for the emission lines used.  
   
\section{Discussion and Conclusions}
Full numerical solutions giving the $Emd_{\rm t}(T_{\rm e})$ and 
$Emd_{\rm app}(T_{\rm e})$ of 
$\epsilon$~Eri were carried out by Ness \& Jordan (2008). The solutions for
different values of $T_{\rm c}$ and fixed $T_{\rm o}$ and $g(r_{c})$ showed 
scaling laws between these parameters and $P_{\rm o}$, $P_{\rm c}$, 
$Em_{\rm t}(T_{\rm o})$ and $Em_{\rm t}(T_{\rm c})$. These solutions were
calculated in spherical symmetry and in hydrostatic equilibrium, including a 
non-thermal pressure term. The link between the scaling laws and the assumed 
energy
balance was not obvious. Here, analytical solutions for a plane parallel 
atmosphere, with some approximations for the variation of $P_{\rm e}$ are
presented, which give scaling laws that can be linked directly to the energy
 balance adopted. These analytical solutions reproduce well the values of
$P_{\rm o}$ and $P_{\rm c}$ and give values of $Em_{\rm t}(T_{\rm o})$ and 
$Em_{\rm t}(T_{\rm c})$ that are smaller than in the full solutions by less 
than about a factor of 1.3. The analytical solutions are therefore useful in
finding the best initial conditions in full numerical solutions. Only the 
values of $g_{*}$, $T_{\rm o}$ and $T_{\rm c}$ are required, the latter two
being simple to measure. 

When a `critical solution' is found by setting the conductive flux
at $T_{\rm o}$ to zero, and when $T_{\rm o}$ is very much less than 
$T_{\rm c}$, our value of $P_{\rm o}({\rm crit})$ is the same as
that given by Hearn's (1975, 1977) minimum energy loss hypothesis.
In our approach, the use of the condition that $Emd_{\rm t}(T_{\rm e})$ passes
through a minimum at some observed $T_{\rm o}$ has the advantage that 
$P_{\rm c}$ and $Em_{\rm t}(T_{\rm o})$ can be found, as well as $P_{\rm o}$ 
and $Em_{\rm t}(T_{\rm c})$. 

We have applied our results to Procyon, for which the approximations made 
should be less accurate than for $\epsilon$~Eri. It is difficult to measure 
the electron density in Procyon, but our results rule out suggestions that 
in the spatially averaged atmosphere the values significantly exceed about 
$7 \times 10^9$~cm$^{-3}$ in the transition region above $10^5$~K, or about 
$3 \times 10^8$~cm$^{-3}$ in the quiescent corona. 
 Improvements to the density-sensitive line ratios in the {\it EUVE}
 wavelength range will have to await the flight of new instruments.
    
Some earlier problems in reconciling coronal and transition region
pressures have been reduced by using the full 
$Emd_{\rm app}(T_{\rm e})$, rather than single $T_{\rm c}$ fits to earlier 
measurements of X-ray fluxes. If the higher electron densities found from
the {\it EUVE} and possibly from the LETGS could be confirmed, these would 
show that active
 region material is indeed present (Schmitt et al. 1985; Jordan et al. 1986).  
  
\section*{Acknowledgments}
~\\
We are grateful to the referee, J. L. Linsky, for his useful comments.\\  
~\\
{\bf REFERENCES} \\
~\\
Brooks D. H., Warren H. P., Williams D. R., \\
\indent Watanabe T., 2009, ApJ, 705, 1522 \\
\noindent
Dere K. P., Landi E., Young P. R., Del Zanna G.,  \\
\indent Landini M., Mason H. E., 2009, A\&A, 498, 915      \\
\noindent 
Drake J. J., Laming J. M., Widing K. G., 1995, ApJ, 443, \\
\indent 393 \\
\noindent
Dupree A. K., Brickhouse N. S., Doschek G. A., Green J. \\
\indent C., Raymond J. C., 1993, ApJ, 418, L41  \\
\noindent
Gabriel A. H., Jordan C., 1969, MNRAS, 145, 241 \\
\noindent
Gibson S. E., Fludra A., Bagenal F., Biesecker D., Del \\
\indent Zanna G., Bromage B. J. I., 1999, J. Geophys. Res., \\
\indent 104, 9691 \\
\noindent
Girard T. M. et al., 2000, AJ, 119, 2428 \\
\noindent
Griffiths N. W., Jordan C, 1998, ApJ, 497, 883 \\
\noindent
Hearn A. G., 1975, A\&A, 40, 355 \\
\noindent
Hearn A. G., 1977, Sol. Phys., 51, 159 \\
\noindent
Jeffrey A., 1995, Handbook of Mathematical Formulas and \\
\indent Integrals, Academic Press, London, p. 147 \\ 
\noindent
Jordan C., 2000, Plasma Phys. \& Controlled Fusion, 42,  \\
\indent 415 \\
\noindent
Jordan C., Brown A., 1981, in Bonnet R.M., Dupree A.K., \\
\indent eds, Solar Phenomena in Stars and Stellar Systems, \\
\indent NATO ASIC, 68, Reidel, Dordrecht, p. 199  \\
\noindent
Jordan C., Brown A., Walter F. M., Linsky J. L., 1986, \\
\indent MNRAS, 218, 465 \\
\noindent
Jordan C., McMurry A.D., Sim S.A., Arulvel M., 2001, \\
\indent MNRAS, 322, L5 \\
\noindent
Jordan C., Ayres T. R., Brown A., Linsky J. L., Simon T., \\
\indent 1987, MNRAS, 225, 903 \\
\noindent
Kopp R.A., 1972, Sol. Phys., 27, 373 \\
\noindent
Landi E., Feldman U., Doschek G. A., 2006, ApJ, 643, 1258 \\
\noindent
Landi E., Del Zanna G., Young P. R., Dere K. P., Mason \\ 
\indent H. E., Landini M., 2006, ApJS, 162, 261 \\ 
\noindent
Liang G. Y., Zhao G., Shi J. R., 2006, MNRAS, 368, 196 \\
\noindent
Maran S. P. et al., 1994, ApJ, 421, 800 \\
\noindent
Mozurkewich D. et al., 1991, AJ, 101, 2207 \\
\noindent
Mullan D. J., Cheng Q. Q., 1994, ApJ, 435, 435 \\
\noindent
Ness J.-U., Jordan C., 2008, MNRAS, 385, 1691  \\
\noindent
Ness J.-U., Mewe R., Schmitt J. H. M. M., Raassen A. J.\\
\indent J., Porquet D., Kaastra J. S., van der Meer R. L. J.,\\
\indent Burwitz V., Predehl P., 2001, A\&A, 367, 282 \\
\noindent 
Neupert W., 1965, AnAp, 28, 446 \\
\noindent
Pan H. C., Jordan C., 1995, MNRAS, 272, 11 \\
\noindent
Philippides D., 1996, D.Phil. Thesis, University of Oxford \\
\noindent
Raassen, A. J. J., Mewe R., Audard M., G\"udel M., Behar \\
\indent E., Kaastra J. S., van der Meer, R. L., Foley C. R., \\
\indent Ness J.-U., 2002, A\&A, 389, 228 \\  
Redfield S., Ayres T.R., Linsky J.L., Ake T.B., Dupree \\
\indent A.K., Robinson R. D., Young P. R., 2003, ApJ, 585,\\
\indent  993 \\
\noindent
Rosner R., Tucker W. H., Vaiana G. S., 1978, ApJ, 220, 643 \\
\noindent 
Sanz-Forcada J., Brickhouse N. S., Dupree A. K., 2003, \\
\indent ApJS, 145, 147 \\
Sanz-Forcada J., Favata F., Micela G., 2004, A\&A, 416, 281 \\ 
\noindent
Schmitt J. H. M. M., Harnden F. R. Jr., Peres G., Rosner \\
\indent R., Serio S., 1985, ApJ, 288, 751 \\
\noindent
Schmitt J. H. M. M., Drake J. J., Haisch B. M., Stern R. \\
\indent A., 1996, ApJ, 467, 841. \\
\noindent
Sim S. A., 2002, D.Phil. Thesis, University of Oxford \\
\noindent
Sim S. A., Jordan C., 2003a, MNRAS, 346, 846 \\
\noindent 
Sim S. A., Jordan C., 2003b, MNRAS, 341, 517 \\
\noindent 
Spitzer L. (Jr.), 1956, {\it Physics of fully ionized gases}, \\
\indent Interscience Publications, Inc, New York \\
\noindent
Torricelli-Ciamponi, G., Einaudi G., Chiuderi C., 1982, \\
\indent  A\&A, 105, L1 \\
\noindent 
Wood B. E., Harper G. M., Linsky J. L., Dempsey R. C., \\ 
\indent 1996, ApJ, 458, 761 \\
 
\label{9}

\end{document}